 \documentclass[pmlr,twocolumn,10pt]{jmlr} 

\usepackage{multirow}
\usepackage{booktabs}
\usepackage{siunitx}

\usepackage[switch]{lineno}

\theorembodyfont{\upshape}
\theoremheaderfont{\scshape}
\theorempostheader{:}
\theoremsep{\newline}

 \title{RELATE: Relation Extraction in Biomedical Abstracts with LLMs and Ontology Constraints}

\author{%
\Name{Olawumi Olasunkanmi}
\Email{olawumi@cs.unc.edu}\\
\addr Department of Computer Science, University of North Carolina, Chapel Hill, United States
\AND
\Name{Matthew Satusky}
\Email{msatusky@renci.org}\\
\Name{Hong Yi} \Email{hongyi@renci.org}\\
\Name{Chris Bizon} \Email{cbizon@renci.org}\\
\addr Renaissance Computing Institute, Chapel Hill, United States
\AND
\Name{Harlin Lee} \Email{harlin@unc.edu}\\
\Name{Stanley Ahalt} \Email{ahalt@unc.edu}\\
\addr School of Data Science and Society, University of North Carolina, Chapel Hill, United States
}

\begin{document}

\maketitle

\begin{abstract}
Biomedical knowledge graphs (KGs) are vital for drug discovery and clinical decision support but remain incomplete. Large language models (LLMs) excel at extracting biomedical relations, yet their outputs lack standardization and alignment with ontologies, limiting KG integration with free texts.
We introduce RELATE, a three-stage pipeline that maps LLM-extracted relations to standardized ontology predicates, e.g., the Biolink Model. The pipeline includes: (1) ontology preprocessing with predicate embeddings, (2) similarity-based retrieval enhanced with SapBERT, and (3) LLM-based reranking with explicit negation handling. This approach performs relation extraction from free-text outputs to structured, ontology-constrained representations.
On the ChemProt benchmark, RELATE achieves 52\% exact match and 94\% accuracy@10, and in 2,400 HEAL Project abstracts, it effectively rejects irrelevant associations (0.4\%) and identifies negated assertions. RELATE captures nuanced biomedical relationships while ensuring quality for KG augmentation.
By combining vector search with contextual LLM reasoning, RELATE provides a scalable, semantically accurate framework for converting unstructured biomedical literature into standardized KGs.
\end{abstract}

\paragraph*{Data and Code Availability}
Data used in this project are publicly available ChemProt dataset\footnote{\url{https://huggingface.co/datasets/bigbio/chemprot}} and PubMed abstracts available through the NIH Helping to End Addiction Long-term (HEAL) Initiative\footnote{\url{https://heal.nih.gov/}}. The code is available in a public Git repository\footnote{\url{https://github.com/RENCI-NER/pred-mapping/releases/tag/v1.0}}.

\paragraph*{Institutional Review Board (IRB)}
This research analyzes public biomedical publication texts on PubMed and does not require IRB approval.

\section{Introduction}
\label{sec:intro}
Knowledge graphs (KGs) serve as structured repositories of facts, capturing entities and their relationships to support applications such as semantic search, recommendation systems, and drug repurposing \citep{olasunkanmi2024explainable,bizon2019robokop}. However, real-world KGs are often incomplete, with missing facts that limit their utility. 

Unstructured biomedical text offers a vast, underutilized source for KG completion. Natural language processing techniques, particularly \textit{relation extraction (RE)}, play a central role in mining relationships between entities mentioned in the text. While Named Entity Recognition (NER) identifies entities (e.g., drugs, proteins, diseases), RE captures the \textit{predicates} describing their relationships (e.g., treats, inhibits), which provide structure and meaning to extracted knowledge. Predicates thus connect entities into an interconnected biomedical KGs. Biomedical texts, however, are challenging to analyze due to domain-specific jargon, abbreviations, and subtle contextual nuances.

Recent advancements in large language models (LLMs) have demonstrated remarkable capabilities in biomedical text understanding \citep{jahan2024comprehensive}. Nevertheless, current LLM-based RE methods often lack integration with domain ontologies, leading to high false-positive rates and inconsistent relationship representations. Standardized \textit{ontology} frameworks such as the Biolink Model \citep{unni2022biolink} define well-structured predicates for biomedical relationships \citep{bizon2019robokop, joachimiak2021kg, reese2021kg, alliance2022harmonizing}. Yet, mapping free-text relations to ontology predicates remains a difficult task due to the semantic gap between natural language expressions and formal ontological representations.

To address these challenges, we propose \textbf{RELATE}, a three-stage pipeline that maps free-form biomedical relations to standardized ontology predicates. The key contributions of this work are:

\begin{enumerate}
    \item \textbf{Ontology-constrained relation standardization:} RELATE introduces the first systematic framework for mapping LLM-extracted biomedical relations to established ontologies (ChemProt and Biolink), moving beyond relation classification toward ontology-grounded standardization. This design ensures semantic interoperability with biomedical KGs. While ChemProt has been used for supervised relation classification \citep{peng2019transfer, warikoo2021lbert}, and Biolink for KG schema definition \citep{joachimiak2021kg, reese2021kg}, our work represents the first application of these ontologies as \emph{standardization targets} for free-text relation extraction, bridging the gap between unstructured text mining and structured knowledge representation.
    
    \item \textbf{Hybrid retrieval and efficient refinement:} We develop a three-stage pipeline that combines ontology-driven preprocessing, SapBERT-enhanced hybrid vector retrieval, and LLM-based contextual reranking. RELATE leverages SapBERT for relation semantics for the first time, and applies computationally intensive reasoning only to small candidate sets ($k=10$), achieving both scalability and accuracy.
    
    \item \textbf{Comprehensive evaluation:} We conduct dual-ontology assessment on ChemProt \citep{peng2019transfer} benchmark and 2,400 real-world HEAL Project abstracts. Integration with an established, large-scale biomedical KG (ROBOKOP) is also planned to further demonstrate its translational impact.
\end{enumerate}

These capabilities enable downstream applications in literature-based discovery, biomedical KG construction, and multi-source knowledge integration; see Appendix~\ref{apd:applications}.

\begin{figure*}[htp]
\floatconts
  {fig:workflow}
  {\caption{The three-stage RELATE pipeline: (1) ontology preprocessing, (2) similarity-based retrieval, and (3) contextual refinement. Ontology preprocessing generates embeddings for predicates and their negated variants. It updates only if the ontology schema or embedding models change. Given an input quadruple—subject, object, relation text, and abstract context—RELATE performs similarity-based retrieval by embedding the relation and retrieving top-$k$ ontology candidates. In the final stage, contextual refinement reranks these candidates with an LLM using the full abstract context, producing the most semantically appropriate ontology predicate.}}
{\centering\includegraphics[width=\textwidth]{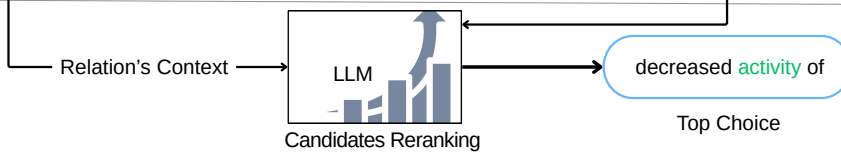}}
\end{figure*}

\section{Related Work}
No existing framework systematically bridges free-text biomedical relation extraction (RE) with standardized ontological predicates. While zero-shot LLMs \citep{li2025grag} can outperform fine-tuned models on smaller datasets, they struggle when confronted with complex, domain-specific tasks that demand specialized knowledge \citep{chen2024benchmarking, jahan2024comprehensive}. Further research has explored adaptive instruction-rich prompting \citep{zhou2024leap, tao2024graphical} and soft prompt-based learning approaches that allow models to automatically optimize prompts rather than relying on hand-crafted instructions \citep{peng2024model}. These methods show promise but have not achieved systematic biomedical ontology integration.

Fine-tuning approaches have demonstrated strong performance on domain-specific benchmarks. Techniques such as adaptive document mapping, ensemble learning with attention mechanisms, and specialized training procedures \citep{shang2025biomedical, zhong2024leveraging, orlova2025challenges} improve classification accuracy. However, they require extensive retraining and do not address the core challenge of standardizing predicates extracted from free-text relations. \citet{chandak2023building} emphasized the importance of standardized schemas and quality control in its PrimeKG framework by integrating heterogeneous data sources. RELATE applies these principles to the RE layer, enabling automated standardization of literature-derived relations for seamless integration into KGs such as PrimeKG, with ensured schema compliance and semantic interoperability.

Recent work in LLM-based RE has explored various validation approaches. RelCheck \citep{ourekouch2025relcheck} addresses low-confidence predictions from pretrained models using ontology-guided LLM validation, though it focuses on confidence rather than standardization. SPIREX \citep{cavalleri2024spirex} applies schema-constrained prompts with graph machine learning validation but restricts itself to RNA-specific domains. 
Graph-augmented approaches include RAG-enhanced LLMs for automated ontology extension \citep{georgakopoulos2025text}, structure-oriented RAG for KG-LLM co-learning \citep{yang2024knowledge}, integrated GNNs with LLM-generated support documents \citep{dong2024graph}, and LLM-driven ontology enrichment pipelines \citep{kollapally2025ontology}. However, these approaches primarily target domain-specific extraction or confidence validation rather than systematic standardization of extracted relationships to established ontological frameworks. Multi-channel neural architectures combine textual and graph structures but focus primarily on structural features rather than systematic predicate mapping. \citet{caufield2024structured} used structured prompts for ontology-grounded annotations, focusing on entity linking rather than full relation standardization, while enhancements to SemRep \citep{ming2024enhancing} remain within existing predicate frameworks.

While \citet{cao2024automatic} built an end-to-end system for extracting knowledge about rare diseases, their work addresses only 6 relation types in a narrow domain. More fundamentally, no existing framework systematically bridges free-text biomedical RE with standardized ontological predicates. RELATE addresses this critical gap by mapping free-text relations to standardized ontology predicates, handling both narrow (ChemProt's 9 predicates) and broad (Biolink's 284 predicates) ontologies, ensuring semantic interoperability and enabling seamless integration into biomedical KGs.

\section{RELATE Methodology}
Biomedical relation extraction (RE) systems face a critical challenge: while LLMs excel at identifying subject–object relationships in scientific text, the extracted relations often lack standardization and fail to align with biomedical ontologies. This disconnect limits their value for KG construction and augmentation. 

To address this limitation, we propose a three-stage pipeline (Figure~\ref{fig:workflow}) that bridges free-text RE with standardized predicates through an ontology-driven protocol. The pipeline consists of: (1) \textbf{Ontology Preprocessing}, which generates positive and negative descriptors for predicates and embeds them into a searchable space; (2) \textbf{Similarity-Based Retrieval}, which efficiently narrows candidate predicates using vector search enhanced with biomedical embeddings; and (3) \textbf{Contextual Refinement}, which leverages LLM reasoning and abstract-level context to rerank candidates and select the most appropriate predicate.

\subsection{Stage 1: Ontology Preprocessing}\label{stages:stages1}
The first stage of our pipeline (Alg.~\ref{alg:preprocessing_pipeline}) extracts ontology predicates with positive and negative descriptors to encode domain knowledge, and transforms them into a searchable embedding database suitable for relation mapping. 

We begin by collecting all predicate definitions from biomedical ontologies such as the Biolink Model toolkit \citep{unni2022biolink} or ChemProt \citep{DBLP:journals/biodb/LiSJSWLDMWL16}, thereby constructing a base vocabulary of biomedical relationships. We will denote the set of all ontology predicates as  $\mathcal{R}$.  Each predicate $r \in \mathcal{R}$ represents a distinct biological or medical relationship (e.g., affects, inhibits) that can exist between entities in the KG.
 
The descriptor dataset $\mathcal{D}_r^+$ is constructed by gathering existing textual descriptions for each predicate $r$. For example, one descriptor for the predicate \textit{affects} may be ``describes an entity that has an effect on the state or quality of another existing entity.'' These descriptions are sourced from multiple authoritative biomedical sources. In the Biolink example, this includes official documentation and standardized vocabularies, while in the ChemProt case, online relation category descriptions suffice. The distributions of the predicate descriptors are presented in Tables \ref{tab:top_predicates_distribution_chemprot} and \ref{tab:top_predicates_distribution_heal}. These aggregated descriptions form the positive dataset $\mathcal{D}_r^+$ and capture the semantic diversity with which biomedical relationships appear in scientific literature.
     
To complement the positive dataset, we introduce a novel strategy for generating negative descriptors. For each descriptor in $D_r^+$, an LLM produces a natural, semantically coherent negation using a structured prompt (Figure~\ref{fig:negation_prompt}). For example, the earlier descriptor becomes ``describes an entity that \underline{does not have} an effect on the state or quality of another existing entity.'' These negated descriptors are stored in a separate dictionary $D_r^-$ to serve as contrastive examples that help the embedding model distinguish valid from invalid predicate mappings. $D_r^-$ is labeled by $r$ + ``\_NEG'' (i.e. \textit{affects} $\to$ \textit{affects\_NEG}) to maintain clear separation from positive instances.

Finally, every \textit{descriptor} in $D^+_r$ and $ D^-_r $ is transformed into a $d$-dimensional vector representation using the embedding model $f_{\theta}$. Instead of aggregating these descriptor embeddings to a single predicate embedding, we store all descriptor embeddings, $\mathcal{V}_r$, in a searchable database for the next stage.

To address limitations of general-purpose embedding models (e.g., \texttt{nomic-text-embed} or \texttt{bge-m3}) in biomedical domains, we introduce an (optional) hybrid retrieval enhancement based on SapBERT \citep{liu2021self}, a model specialized in biomedical entity linking. In this setup (Figure \ref{fig:workflowplus}), SapBERT was fine-tuned on $D^+_r$ and $ D^-_r $ to generate a second set of embeddings $\mathcal{V}'_r$. While SapBERT has demonstrated effectiveness in biomedical entity alignment, this work presents its first application to RE through descriptor embedding enhancement. Our hybrid approach leverages SapBERT's biomedical domain knowledge to improve relation semantic similarity beyond general-purpose embedding models.

\begin{figure}[t]
\floatconts 
{fig:negation_prompt}
{\caption{Negation generation prompt used in Stage 1 (Section~\ref{stages:stages1}) for creating negative descriptors from positive descriptors.}}
{\setlength{\fboxsep}{0pt}
\fbox{\begin{minipage}{0.5\textwidth}
\small\texttt{You are a biomedical researcher extracting negations of ontological predicates.\\
\\
Your Task:\\
Given a description, return its natural negation.\\
\\
Rules:\\
1. Preserve the meaning but negate the entire description.\\
3. Do not summarize or change the structure of the descriptor text.\\
4. If there is not enough information to create a negation, your response should be "NOT ENOUGH INFORMATION"\\
4. Only return the negation—no explanations or extra text.\\
\\
Examples:\\
- "has effect" → "does not have effect"\\
- "during which ends" → "during which does not ends"\\
- "happens during" → "does not happen during"\\
\\
Input: "\{descriptor\_text\}"\\
\\
Output: A JSON object with these exact keys and format:\\
\{\{"negation\_of\_the\_descriptor\_text": "negated version" or "NOT ENOUGH INFORMATION"\}\}}
\end{minipage}}}
\end{figure}

\subsection{Stage 2: Similarity-Based Retrieval}\label{stages:stages2}
For every predicate $r \in \mathcal{R}$, Stage 1 has produced positive and negative descriptor dictionaries $\mathcal{D}_r^+$, $\mathcal{D}_r^-$ and corresponding descriptor embedding vectors $\mathcal{V}_r$. We now leverage these data to retrieve candidate ontology predicates for incoming relations extracted by an upstream RE system, which could be any general-purpose LLM. 

The extracted relations come in a quadruple $(s, o, T, a)$: $s$ subject entity, $o$ object entity, $T$ the free-form relation text, and $a$ the abstract context from which the relation was identified. For example, in Figure \ref{fig:workflow}, $T$ is ``decreases transporting activities of,''  which does not exist in Biolink ontology $\mathcal{R}$ in this format. 

First, free-form relation text $T$ is embedded to a query vector $q=f_\theta(T)$. We use the same model $f_{\theta}$ from Stage~1, ensuring semantic consistency between query relations and stored ontology predicates.

Once embedded, the query vector $q$ is compared against all predicate descriptor embeddings in $\mathcal{V}_r,\; ~\forall r \in \mathcal{R}$. This similarity search retrieves the top-$k$ candidates (typically $k=10$) using cosine similarity. If multiple descriptors of the same predicate are in the top-$k$ results, they are collapsed into a single predicate candidate. When SapBERT enhancement is on, the top-$k$ candidates in $\mathcal{V}'_r,\; ~\forall r \in \mathcal{R}$ are retrieved in parallel. Then, the two sets of candidates are merged and de-duplicated to identify (up to) 2$k$ candidate predicates. 

By narrowing the search space from hundreds of ontology predicates to a small candidate set, this stage provides an efficient yet semantically informed mechanism for predicate selection. At this point, however, the search operates without the benefit of contextual information from the original biomedical abstract $a$, focusing only on lexical and semantic similarity.

\subsection{Stage 3: Contextual Refinement} \label{stages:stages3}

\begin{figure}[h]
\floatconts{fig:reranking_prompt}
{\caption{Contextual reranking prompt used in Stage 3 (Section~\ref{stages:stages3}) for LLM-based predicate selection with explicit negation handling.}}
{\setlength{\fboxsep}{0pt}
\fbox{\begin{minipage}{0.5\textwidth}
\small\texttt{You are an expert in biomedical relationships. Based on the text below:\\
\\
Subject: \{subject\}\\
Object: \{object\}\\
Original Relationship: \{relationship\}\\
Abstract: \{abstract\}\\
\\
Candidate Predicates:\\
\{choices\_str\}\\
\\
Instructions:\\
- Choose the best predicate from the list that matches the intended meaning and direction.\\
- If the original relationship implies negation (e.g., "does not cause"), select the matching base predicate, but set "negated" to "True".\\
- If no match exists, return `"mapped\_predicate": "none"'.\\
\\
Respond with ONLY this JSON object:\\
\{\{"mapped\_predicate": "one of the predicate keys or `none'", "negated": "True" or "False"\}\}}
\end{minipage}}}
\end{figure}

The final stage addresses the key limitation of similarity-based retrieval: its reliance on context-free embeddings. While Stage~2 efficiently narrows the predicate space to a small candidate set, it does not incorporate the biomedical context in which the relation was expressed. Stage~3 introduces contextual refinement, where an LLM evaluates and reranks the top-$k$ candidates against the full relation context.

The LLM is given structured prompt (Figure~\ref{fig:reranking_prompt}) along with the following information: the top candidate predicates retrieved by similarity-based search and the full RE quadruple $(s, o, T, a)$ including the abstract context. By explicitly grounding the candidate predicates in this richer context, the LLM can better assess which ontology predicate is most appropriate for the specific biomedical scenario.

The LLM then performs reranking of the candidate set, selecting the single predicate that best aligns with the domain context. Importantly, the system includes a \texttt{NONE} option, allowing the model to reject all candidates if none provide a semantically valid match. This safeguard prevents the forced assignment of predicates in cases where the extracted relation is either too domain-specific, erroneous, or incompatible with the ontology. 

The output of Stage~3 is a contextually validated predicate $r^*$ that integrates both semantic similarity and biomedical context. This refinement ensures that only relationships consistent with both the ontology and the source text are admitted into the KG, thereby improving accuracy and reducing false positives.

\section{Experiments}
\label{sec:experimental_setup}

We assess RELATE on two independent datasets with different ontology models, as well as four different embedding model configurations. See Appendix~\ref{apd:data_descriptions} for example descriptors in both datasets. 

\subsection{Datasets}
\label{sec:evaluation_framework}

The ChemProt dataset \citep{DBLP:journals/biodb/LiSJSWLDMWL16} provides annotated chemical-protein relations representing diverse pharmacological and biochemical relationship types, including regulators, modulators, agonists, antagonists, and substrate interactions. We treat the information in ChemProt as ground truth abstract-predicate pairs for evaluation purposes. Standard metrics in information retrieval are used, such as exact match accuracy, accuracy@$k$, and Mean Reciprocal Rank (MRR). Recall from Figure~\ref{fig:reranking_prompt} that Stage 3 of RELATE only outputs its top choice predicate $r^*$ or \texttt{NONE}, and does not return a fully reranked or reordered list of $k$ candidates. Therefore, exact match accuracy is calculated with respect to the final output of RELATE ($r^*$), but accuracy@$k$ and MRR are calculated using the top-$k$ candidates after Stage 2, before LLM-based Stage 3. This allows us to analyze the impact of LLM-based refinement in isolation.

Next, we apply RELATE to 2,400 PubMed abstracts from the HEAL Project\footnote{\url{https://heal.nih.gov}} targeting opioid use disorder research. This real-world dataset reflects the practical scenario in which extracted relationships must be mapped to standardized Biolink predicates for integration into a large biomedical KG\footnote{\url{https://robokop.renci.org/}}. Unlike ChemProt, the HEAL abstracts lack ground-truth annotations, requiring qualitative assessment of RELATE's outputs. 

Importantly, this corpus captures the complexity and noisiness of real biomedical literature. In contrast to benchmark datasets, these abstracts often contain methodological statements or negative scientific findings (e.g., ``drug does not affect condition,'' ``gene not associated with disease''). Such cases underscore the necessity of rejection and negation-handling to ensure only semantically appropriate biomedical relationships enter the KG. 

\subsection{LLM and Embedding Models}
\label{sec:model_configurations}
We experimented with MedGemma (4B), LLaMA2 (7B), LLaMA3 (8B), LLaMA3 (70B), MedGemma (27B), and OpenAI GPT-4o-mini (unknown number of parameters). The last two demonstrated greater consistency and fewer hallucinations compared to the other models. Since those two models had very similar performances, we only report results using MedGemma (27B) in this work. However, note that any other LLM could be substituted to reproduce the pipeline. MedGemma (27B) is Google's medical language model \citep{sellergren2025medgemma}, trained exclusively on medical text and optimized for efficient inference through architectural improvements in performance and computational efficiency. 

To evaluate RELATE's pipeline contribution beyond raw LLM capability, we implemented a direct classification baseline. This baseline uses MedGemma (27B) to classify ChemProt relations by providing the same input quadruple (subject, object, relation text, abstract) along with the list of 9 ChemProt predicates. This represents a common approach in LLM-based relation extraction and provides a fair comparison since it uses the same LLM and has access to the same predicate definitions.

As with the LLMs, we also experimented with several embedding approaches. While we prioritize accessibility through \texttt{ollama}-based open-access models, we tested OpenAI's \texttt{text-embed-large} models and others with 8{,}192-token context windows that provide general-purpose semantic understanding. These include \texttt{nomic-embed-text} \citep{nussbaumnomic}, a 768-dimensional embedding model, and \texttt{bge-m3}, a 1{,}024-dimensional multilingual embedding model for diverse textual contexts \citep{chen2024bge}. Since \texttt{nomic-embed-text} and \texttt{bge-m3} achieved comparable performance, we report only the results obtained with the 768-dimensional \texttt{nomic-embed-text} in this work.  

For (optional) hybrid retrieval, we experimented with adding fine-tuned SapBERT~\citep{liu2021self}. 
We initialized the model with \texttt{BiomedNLP-PubMedBERT-base-uncased-abstract} \texttt{-fulltext} and optimized it using a contrastive learning framework. Specifically, we employed multi-similarity loss with hard negative mining, which encourages the model to better distinguish between semantically close but incorrect pairs. To keep training efficient while still capturing relevant semantic features, we limited input sequences to 25 tokens and represented each instance using the \texttt{[CLS]} embedding.

Training was carried out for 10 epochs with a batch size of 256 and a learning rate of $2 \times 10^{-5}$. We used mixed-precision training to reduce computational overhead and set a fixed random seed for reproducibility. 

\begin{table*}[t]
\floatconts
{tab:chemprot_overall}
{\caption{RELATE performance on ChemProt. \textbf{Stage 2 (Candidate Retrieval)}: a@$k$ = accuracy in top-$k$ candidates \textit{before} LLM-based refinement; MRR = mean reciprocal rank. \textbf{Post-Refinement}: Exact Match = final accuracy \textit{after} LLM-based refinement; \textit{Baseline} = direct LLM classification without RELATE's retrieval-refinement pipeline. LLM used is MedGemma (27B) for all.}} 
{\begin{tabular}{ccccccc}
\toprule
\multirow{2}{4cm}{\centering\textbf{Embedding Model}}& \multicolumn{5}{c}{\textbf{Stage 2 (Candidate Retrieval)}} & \textbf{Post-Refinement} \\
\cmidrule(lr){2-6} \cmidrule(lr){7-7}
 & a@1 & a@3 & a@5 & a@10 & MRR & Exact Match (a@1) \\
\midrule
 \texttt{nomic} & \textbf{0.528} & \textbf{0.686} & 0.744 & 0.755 & \textbf{0.612} & 0.464 \\
 \texttt{bge-m3} & 0.468 & \textbf{0.686} & \textbf{0.788} & 0.812 & 0.590 & 0.467 \\
 \texttt{nomic} + SapBERT & 0.100 & 0.328 & 0.688 & 0.914 & 0.300 & \textbf{0.520} \\
\texttt{bge-m3} + SapBERT & 0.100 & 0.321 & 0.691 & \textbf{0.940} & 0.302 & 0.516 \\
Baseline (no embedding) & - & - & - & - & - & 0.220 \\
\bottomrule
\end{tabular}}
\end{table*}

\section{Results and Discussions}\label{sec:results}
\paragraph{RELATE Pipeline is Necessary.}
To validate RELATE's pipeline design, we compared against direct LLM classification where MedGemma (27B) selects from the 9 ChemProt predicates without retrieval-refinement. The baseline achieves only 22.0\% exact match (Table~\ref{tab:chemprot_overall}), compared to RELATE's 52.0\% (136\% improvement). This demonstrates that structured retrieval outperforms asking an LLM to select directly from a predicate list. The baseline's poor performance reflects the difficulty of simultaneous predicate understanding and selection without semantic guidance from embedding-based retrieval.

The exact match accuracy of 52\% reflects the difficulty of mapping to fine-grained ontology predicates. However, this metric may understate practical utility. Manual analysis of the misclassifications show semantically acceptable near-misses, such as mapping to \textit{agonist} instead of \textit{modulator}. When allowing such semantically close alternatives, accuracy increases to approximately 70-80\%.

\paragraph{LLM-based Refinement is Necessary.}

Table \ref{tab:chemprot_overall} illustrates how embedding choice shapes RELATE's performance on the ChemProt benchmark. Baseline configurations with \texttt{nomic} and \texttt{bge-m3} produce stronger top-rank accuracy (a@1 up to 0.528, MRR $\approx$ 0.6). In contrast, SapBERT-enhanced embeddings yield slightly higher Exact Match (0.520) and dramatically improve retrieval coverage, with a@10 reaching 0.940. This indicates that SapBERT ensures the correct ontology predicate is almost always retrieved, though often ranked lower in the list (a@1 drops to 0.1, MRR $\approx$ 0.3).

These results highlight a trade-off: baseline embeddings rank the correct answer more accurately at the top, while SapBERT embeddings broaden coverage and guarantee inclusion within candidate sets. 

While selecting the top-ranked candidate from embeddings without LLM refinement would yield higher exact match on this benchmark, such an approach assumes embedding similarity reliably ranks correct predicates first across all datasets and relation types. This assumption cannot be guaranteed a priori. RELATE's design prioritizes robustness through SapBERT's comprehensive retrieval combined with LLM-based contextual reranking, which serves as a buffer against cases where embedding similarity alone performs poorly. This allows RELATE to recover early precision while preserving high recall, transforming broad retrieval into precise ontology-constrained mappings regardless of how initial embeddings behave on new corpora.

\paragraph{When Does RELATE Reject Candidates?}

RELATE's rejection capability (\texttt{NONE}) represents a critical quality control mechanism. Out of 2,400 PubMed abstracts from the HEAL project, 10 relationships were appropriately rejected as unsuitable for Biolink predicate mapping (rejection rate: ~0.4\%). Manual analysis of these rejections reveals contextual understanding that distinguishes between legitimate biomedical relationships and other types of associations commonly found in scientific literature. 

The rejected relationships fell into distinct categories reflecting non-biomedical associations: genetic nomenclature specifications (3 cases), clinical procedures and documentation (4 cases, including ``Clinicians prescribe Opioid" and ``Women on BUP had documented Opioid misuse"), methodological relationships (2 cases, such as ``Anisomycin used to detect GNPTG"), and specialized reporting contexts (1 case). These rejections demonstrate the pipeline's ability to recognize when extracted relationships represent procedural, methodological, or administrative associations rather than biological or medical relationships suitable for KG representation.

These rejections occurred despite high embedding similarity scores, demonstrating the reranker's comprehensive contextual analysis. The low rejection rate (0.4\%) suggests that the majority of extracted relationships represent valid biomedical associations while maintaining strict quality standards for edge cases requiring specialized handling. All 10 rejections were deemed appropriate (100\% precision).

On ChemProt, there were 5 correct rejections out of 18 total rejections. These false rejections occur where, despite high similarity candidate scores, RELATE failed to produce mappings. Cases like ``loperamide modifies corticotropin-releasing hormone action'' (0.661 score for \textit{modulator}) and ``ammonium sulfate reduces CBG concentrations'' (0.632 for \textit{downregulator}) were rejected despite appropriate candidates. While some discrepancies reflect stricter semantic criteria, others suggest overly conservative contextual reranking, highlighting evaluation challenges with traditional benchmarks.
These contrasting patterns reflect differences in semantic alignment between upstream extraction and ontology predicates across datasets. See more details in Appendix~\ref{appendix:rejections}.

\paragraph{When Does RELATE Choose Negated Candidates?}
Out of 2,400 PubMed abstracts from the HEAL project, 77 relationships were correctly flagged as negated assertions, revealing how well the system processes the nuanced ways of reporting biomedical negative findings. Most prevalent was ``not associated with", which accounted for nearly half of the negated cases (32 instances). Beyond these straightforward cases, we found the pipeline successfully navigating more complex terrain. Other negations like ``does not affect" appeared in 8 cases, while direct causal negations (``does not cause") showed up 7 times.
The remaining quarter of cases presented genuine linguistic challenges. Consider phrases like ``remained impaired despite treatment" or ``not necessary for acute insulin-induced uptake." These require contextual understanding that goes well beyond pattern matching. The system had to recognize that ``fails to affect" and ``unable to deliver" carry negative semantic weight, even when the word ``not" never appears. See more details in Appendix~\ref{apd:negation}.

Perhaps most importantly, RELATE maintained conceptual accuracy throughout this process. When it encountered ``Drug X does not treat Disease Y," it correctly identified the underlying relationship as therapeutic (mapping to \textit{treats}) while flagging the negation. This dual parsing of both the relationship type and its polarity proved consistent across our dataset. The most frequent base relationships were association patterns (28 cases) and effect relationships (18 cases), followed by causal relationships (8 cases).
The other application capability of negation handling includes findings on use, such as which treatments fail, which genetic variants are not linked to diseases, and which interventions prove ineffective.

\paragraph{SAPBERT Enhancement Helps.}  
To evaluate SapBERT's effect, we compared 2,139 predicate assignments between \texttt{bge-m3} and \texttt{bge-m3 + SapBERT}.
Their disagreement patterns reveal four key improvements by SAPBERT aligned with biomedical expert judgment. Clinical terminology refinement was most prominent, with 28 cases shifting from \textit{associated with: increased likelihood of} to \textit{predisposes to condition}. Domain specificity enhancement appeared in 12 cases where generic \textit{associated with} became \textit{gene associated with condition} for relationships like ``DRD4 associated with Heroin Addiction," recognizing that genetic contexts require more informative predicates.

Evidence-based caution moderated 12 strong causal claims, changing \textit{causes} to \textit{contributes to} for relationships like ``MDMA causes Depression," reflecting the complex multifactorial causation typical in biomedicine, where direct causality claims often overstate evidence strength. Mechanistic precision enhanced molecular relationship descriptions, with 10 cases shifting from \textit{affects: increased abundance} to \textit{affects: increased activity or abundance} for relationships like ``Betaine enhances phosphorylation of STAT3," capturing both quantitative and functional effects in biochemical interactions. These systematic patterns demonstrate that SapBERT's biomedical entity embeddings guide predicate selection toward clinically appropriate, evidence-aware terminology that better reflects expert knowledge of biomedical relationship complexity.

\paragraph{Limitations.} RELATE's performance depends on the quality of upstream entity and relation extraction. While we do not assume perfect extraction, the input quadruples must be reasonably accurate. Errors in entity boundaries, entity type misclassifications, or poorly extracted relation phrases will propagate through the standardization pipeline. In production KG construction, additional methods address these entity errors. RELATE's modular design allows it to benefit from improvements in upstream extraction systems without requiring system redesign.

\section{Conclusion}\label{sec:conclude}
We introduced RELATE, a three-stage pipeline for automated ontology-constrained predicate mapping in biomedical relation extraction. By combining ontology-driven preprocessing, efficient vector search, and context-aware LLM-based reranking, RELATE systematically converts free-text relations into ontology-aligned and interoperable KG edges. Experiments on ChemProt demonstrate the trade-offs between early precision and retrieval depth, while application to 2,400 HEAL Project abstracts illustrates RELATE's ability to handle noisy real-world biomedical literature, including detecting negated assertions and appropriately rejecting non-biomedical associations. Together, these results establish a foundation for scalable, ontology-constrained relation extraction and highlight a path forward for enriching biomedical KGs with high-quality, standardized information. Future work includes improving computational overhead, incorporating further top-$k$ reordering and evaluation in the refinement stage, designing a scalable evaluation that combines expert judgment with automated consistency checks, and integration into the existing large biomedical KG (ROBOKOP). As part of the integration plan, the HEAL abstract authors and domain experts are being recruited to systematically evaluate the quality and reliability of RELATE and the LLM-extracted quadruples that serve as its inputs.

\acks{We thank the LitCoin project partners for their immense contributions, and the HEAL initiative consortium for providing access to opioid-related research abstracts. Also, we acknowledge the RENCI's Data Analytics group and Knowledge Graph Working Group for constructive reviews. This research was supported by NIH \#75N95023C00032 (LITCOIN)}.

\bibliography{jmlr}

\appendix

\section{Clinical and Research Applications}\label{apd:applications}

\subsection{RELATE's Role and Significance}
RELATE is a relation standardization component that ensures relations extracted from biomedical text are accurately mapped to standardized ontology predicates. This standardization is foundational for downstream knowledge graph applications.

In literature-based discovery, standardized predicates enable systematic querying across diverse sources. A researcher investigating drug repurposing can query for all compounds that upregulate proteins downregulated in a specific disease, confident that the \textit{upregulator} predicate maintains consistent semantics across thousands of papers. Without standardization, the same relationship might appear as ``increases expression,'' ``upregulates,'' or ``enhances activity,'' requiring separate queries and potentially missing relevant findings.

Biomedical knowledge graph systems such as ROBOKOP require ontology-aligned relations for reliable reasoning and inference. RELATE ensures that literature-derived evidence integrates seamlessly with structured knowledge bases, enabling complex queries that traverse multiple relationship types and data sources.

Our evaluation on HEAL demonstrates how relations from distinct domains can be mapped to common Biolink predicates, while ChemProt enables unified knowledge graphs that integrate chemical interactions.

\subsection{Comparison with End-to-End Systems}
Unlike end-to-end systems such as AutoRD, which perform extraction, standardization, and graph construction, RELATE focuses specifically on standardization. This provides modularity, allowing RELATE to work with any upstream extraction system, and scalability, handling 284 Biolink predicates compared to 6 relation types in AutoRD.

\subsection{Broader Applicability}
While evaluated on biomedical corpora, RELATE generalizes to any domains with structured predicate vocabularies for computing embeddings.

\section{RELATE Stages Algorithms}\label{apd:first}
Algorithms~\ref{alg:preprocessing_pipeline}, \ref{alg:vector_search}, and \ref{alg:contextual_reranking} summarize the three stages of RELATE. When the optional SAPBERT workflow enhancement is on, RELATE follows the modified pipeline described in Figure~\ref{fig:workflowplus}.
\begin{algorithm2e}
\caption{Ontology Preprocessing}
\label{alg:preprocessing_pipeline}
\KwIn{Biolink Model YAML $\mathcal{B}$, Biolink Toolkit $\mathcal{T}$, Embedding model $f_{\theta}$, LLM}
$\mathcal{R} \leftarrow$ \textsc{ExtractPredicateMappings}$(\mathcal{B}, \mathcal{T})$\;
$\mathcal{D}^+ \leftarrow \emptyset$\;
\For{each predicate $r \in \mathcal{R}$}{
   $\text{descriptions} \leftarrow$ \textsc{GatherDescriptions}$(r)$\;
   $\mathcal{D}^+[r] \leftarrow \text{descriptions}$\;
}
$\mathcal{D}^- \leftarrow \emptyset$\;
\For{each predicate $p \in \mathcal{D}^+$}{
    $\text{neg\_descriptions} \leftarrow \emptyset$\;
    \For{each description $d \in \mathcal{D}^+[p]$}{
        $\text{prompt} \leftarrow$ \textsc{CreateNegationPrompt}$(d)$\;
        $d^{-} \leftarrow LLM(\text{prompt})$\;
        $\text{neg\_descriptions} \leftarrow \text{neg\_descriptions} \cup \{d^{-}\}$\;
    }
    $\text{neg\_key} \leftarrow p + \text{``\_NEG"}$\;
    $\mathcal{D}^-[\text{neg\_key}] \leftarrow \text{neg\_descriptions}$\;
}
$\mathcal{V} \leftarrow \emptyset$\;
\For{each predicate $p \in \mathcal{D}^+ \cup \mathcal{D}^-$}{
    $\text{descriptions} \leftarrow \mathcal{D}^+[p] \text{ if } p \in \mathcal{D}^+ \text{ else } \mathcal{D}^-[p]$\;
    $\text{combined\_text} \leftarrow$ \textsc{CombineDescriptions}$(\text{descriptions})$\;
    $v \leftarrow f_{\theta}(\text{combined\_text})$\;
    $\mathcal{V} \leftarrow \mathcal{V} \cup \{(\text{label}: p, \text{embedding}: v)\}$\;
}
\Return Embedding store $\mathcal{V}$, Positive descriptors $\mathcal{D}^+$, Negative descriptors $\mathcal{D}^-$\;
\end{algorithm2e}
\begin{algorithm2e}
\caption{Similarity-Based Retrieval}
\label{alg:vector_search}
\KwIn{Loose quadruples $\{(s_i, o_i, T_i, a_i)\}$, Predicate embeddings $\mathcal{V}$, Embedding model $f_{\theta}$, Parameter $k$}
\tcp{Relationship Embedding Phase}
$\mathcal{Q} \leftarrow \emptyset$\;
\For{each quadruples $(s_i, o_i, T_i, a_i)$}{
    $q_i \leftarrow f_{\theta}(T_i)$\;
    $\mathcal{Q} \leftarrow \mathcal{Q} \cup \{(i, q_i)\}$\;
}
\tcp{Cosine Similarity Search and Top-k Candidate Retrieval}
$\mathcal{C} \leftarrow \emptyset$\;
\For{each $(i, q_i) \in \mathcal{Q}$}{
    $\text{similarities} \leftarrow \emptyset$\;
    \For{each $(p, v) \in \mathcal{V}$}{
        $\text{sim} \leftarrow \cos(q_i, v)$\;
        $\text{similarities} \leftarrow \text{similarities} \cup \{(p, \text{sim})\}$\;
    }
    $\mathcal{C}[i] \leftarrow$ \textsc{Top-$k$}$(\text{similarities})$\;
}
\Return Candidate sets $\mathcal{C}$\;
\end{algorithm2e}
\begin{algorithm2e}[t]
\caption{Contextual Refinement}
\label{alg:contextual_reranking}
\KwIn{Loose quadruples $\{(s_i, o_i, T_i, a_i)\}$, Candidate sets $\mathcal{C}$, LLM}
$\mathcal{A} \leftarrow \emptyset$\;
\For{each quadruple $(s_i, o_i, T_i, a_i)$}{
    $\text{predicate\_list} \leftarrow$ \textsc{ExtractPredicates}$(\mathcal{C}[i])$\;
    $\text{prompt} \leftarrow$ \textsc{CreateRerankingPrompt}($(s_i, o_i, r_i, a_i)$, \text{predicate\_list})
    
    \tcp{LLM Selection and Validation}
    $r_i^* \leftarrow LLM(\text{prompt})$\;
    \If{$r_i^* = \texttt{NONE}$ or $r_i^* \notin \text{predicate\_list}$}{
        $r_i^* \leftarrow \emptyset$\;
    }
    
    $\mathcal{A} \leftarrow \mathcal{A} \cup \{(s_i, o_i, r_i^*, a_i)\}$\;
}
\Return Mapped quadruples $\mathcal{A}=\{(s_i, o_i, r_i^*, a_i)\}$\;
\end{algorithm2e}
\newpage
\begin{figure*}[t]
\floatconts
  {fig:workflowplus}
  {\caption{SapBERT-augmented RELATE pipeline as in Figure \ref{fig:workflow}. The ontology preprocessing now involves both LLM-generated embeddings and SapBERT finetuned embeddings. Additional changes include dual embedding of the relation text, dual similarity searches on the stored embeddings, and merging of the two top-$k$ candidates.
  }}
  {\raggedright\includegraphics[width=0.95\textwidth]{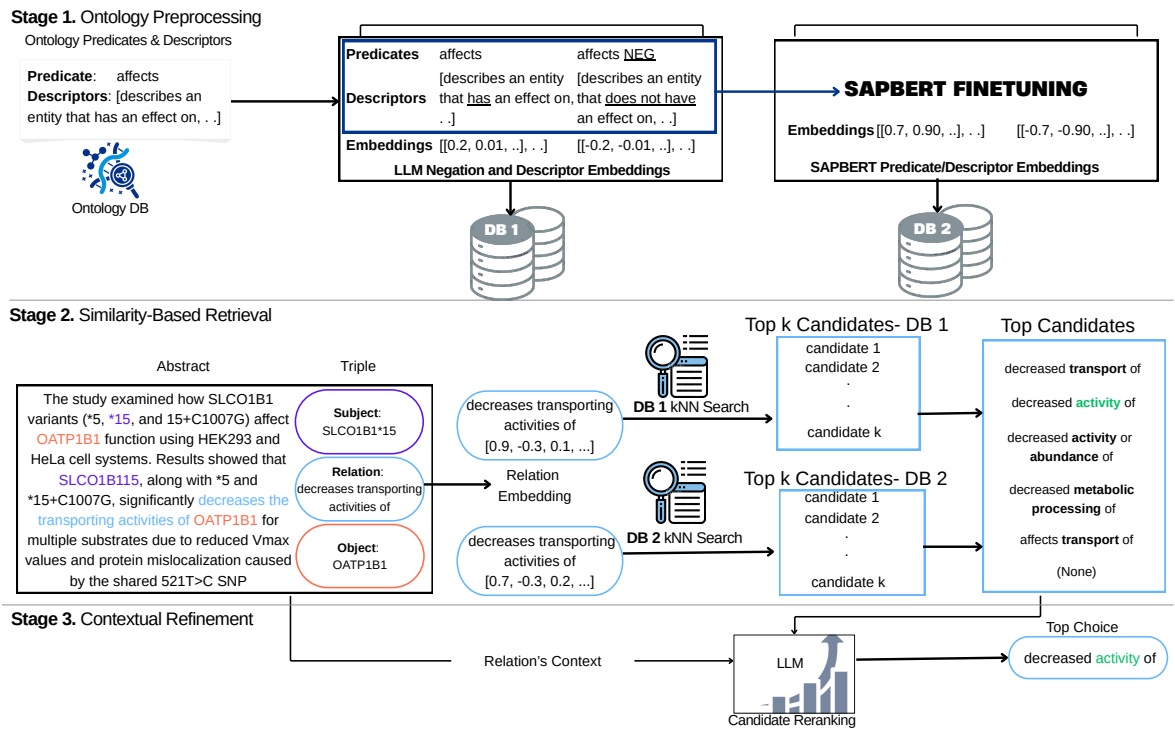}}
\end{figure*}
\newpage
\section{Additional Analysis}
\label{apd:third}

\subsection{ChemProt Pre/Post Refinement Accuracies: Error Attribution Analysis}\label{subsec:error_attribution}

Our results revealed an unexpected pattern: Stage 2 (Section~\ref{stages:stages2}) retrieval achieved 52.8\% accuracy@1, but after Stage 3 (Section~\ref{stages:stages3}) refinement, final accuracy dropped to 46.6\% (Table~\ref{tab:chemprot_overall}). To understand this decline, we analyzed the 6.2\% set failures and attributed them to specific causes.

\subsubsection{Stage 3 Impact Analysis}

We identified all cases where Stage 3 (Section~\ref{stages:stages3}) changed Stage 2's (Section~\ref{stages:stages3}) @1 prediction and categorized them by outcome:

\begin{enumerate}
    \item Stage 3 harmful changes (146 cases): Stage 3 changed a correct @1 prediction to an incorrect answer
    \item Stage 3 beneficial changes (103 cases): Stage 3 rescued failures by selecting the correct answer from ranks 2-10 
\end{enumerate}
    
The net impact was -43 cases (146 harmful - 103 beneficial), accounting for the observed 6.2\% accuracy decline. This analysis focuses on the 146 harmful changes to understand why Stage 3 overrode correct predictions.
\subsubsection{Systematic Over-Specification Pattern}

Detailed examination of the 146 harmful changes revealed highly concentrated error patterns rather than random failures (Table~\ref{tab:stage3_errors}). A single transformation pattern in Stage 2 correctly predicted ``downregulator'' but Stage 3 changed it to ``inhibits''; accounting for 50.7\% of all harmful changes. The top five patterns captured 83.6\% of errors, indicating systematic bias.

\begin{table*}[t]
\centering
\caption{Stage 3 Over-Specification Error Patterns with Original Free-Text Relationship Variants}
\label{tab:stage3_errors}
\footnotesize
\begin{tabular}{llp{6.5cm}rr}
\toprule
\textbf{Ground Truth} & \textbf{Stage 3} & \textbf{Original Free-Text Variants} & \textbf{Count} & \textbf{\%} \\
\midrule
downregulator & inhibits & ``inhibits'' (68), ``inhibits the activity of'' (4), ``inhibits secretion of'' (1), ``inhibits the activities of'' (1) & 74 & 50.7 \\
downregulator & antagonist & ``blocks'' (8), ``inhibits'' (6), ``blockers of'' (1), ``is a poor inhibitor of'' (1), ``tyrosine kinase inhibitor'' (1), ``interferes with inhibitory action on'' (1) & 18 & 12.3 \\
downregulator & inhibitor & ``is an inhibitor of'' (3), ``inhibits'' (3), ``inhibitor of'' (2), ``selectively inhibited'' (1), ``preferentially inhibits'' (1), ``inhibitory activity'' (1), +4 more & 15 & 10.3 \\
upregulator & agonist & ``activates'' (4), ``induces expression of'' (3), ``stimulates'' (2), ``stimulates the transcription of'' (1), ``increases phosphorylation of'' (1) & 11 & 7.5 \\
upregulator & activates & ``activates'' (4) & 4 & 2.7 \\
\midrule
\multicolumn{3}{l}{Top 5 patterns} & 122 & 83.6\% \\
\midrule
upregulator & modulator & ``activates'' (1), ``stimulates activity of'' (1), ``stimulates the activity of'' (1) & 3 & 2.1 \\
downregulator & regulator & ``inhibits'' (1), ``is abolished by'' (1), ``reduces concentrations of'' (1) & 3 & 2.1 \\
downregulator & blocks & ``blocks'' (3) & 3 & 2.1 \\
substrate & regulator & ``impacts metabolism of'' (2) & 2 & 1.4 \\
\multicolumn{3}{l}{Other patterns (6 additional)} & 12 & 8.2 \\
\midrule
\multicolumn{3}{l}{\textbf{Total}} & \textbf{146} & \textbf{100.0\%} \\
\bottomrule
\end{tabular}
\end{table*}

Analysis of the original free-text relationships revealed the fundamental mechanism: 
\begin{enumerate}
    \item Stage 3 preserved mechanistic specificity from source text rather than maintaining ontology-level abstraction. In the dominant error pattern (downregulator → inhibits), 92\% of cases (68/74) had ``inhibits'' as the original free-text relationship
    \item Stage 2 (retrieval-based) correctly abstracted to the ontology-level category ``downregulator,'' matching ChemProt's annotation guidelines. However, Stage 3 (LLM-based refinement) reintroduced the mechanistic specificity from the source text, selecting ``inhibits'' because that was the literal wording in the abstract.
\end{enumerate}

This pattern extended across error types: ``blocks'' appeared in 61\% (11/18) of cases where Stage 3 incorrectly changed ``downregulator'' to ``antagonist,'' and ``activates'' appeared in 73\% (8/11) of cases where Stage 3 changed ``upregulator'' to ``agonist.''

\subsubsection{Interpretation}

These findings illuminate a fundamental tension between retrieval-based and generative approaches to relation extraction. The embedding-based retrieval (Stage 2 (Section~\ref{stages:stages2})) successfully learned ontology-level abstractions from the training data, correctly mapping mechanistic phrases to categorical predicates. However, the LLM refinement (Stage 3 (Section~\ref{stages:stages2})), despite being trained extensively on biomedical literature, demonstrated a preference for preserving the literal mechanistic terminology found in source text.

The concentration of errors (84\%) captured by just five patterns suggests that targeted interventions could recover most of the 6.2\% accuracy loss:

\begin{enumerate}
    \item Confidence-based bypass: Skip Stage 3 when Stage 2's top-1 confidence exceeds a threshold (e.g., 0.8)
    \item Ontology-constrained prompting: Explicitly instruct the LLM to maintain categorical abstraction levels
    \item Restricted candidate sets: Only allow Stage 3 to choose from predicates at the same abstraction level as the top-k Stage 2 candidates
\end{enumerate}

\subsection{Rejected Relations}\label{appendix:rejections}

RELATE includes a rejection mechanism (\texttt{NONE} option, Section~\ref{stages:stages3}) to filter relations that cannot be reliably mapped to ontology predicates. Here, we analyze rejection patterns across both datasets.

\subsubsection{HEAL Dataset Rejections}

In the HEAL dataset of 2,400 abstracts, RELATE rejected 10 relationships (rejection rate: 0.4\%). Manual review confirmed that all 10 rejections were appropriate (100\% precision), filtering non-biomedical associations.

The rejected relationships fell into distinct categories:

\begin{enumerate}
\item Genetic nomenclature specifications (3 cases): Relations describing genetic naming conventions rather than biological relationships (e.g., gene symbol designations)

\item Clinical procedures and documentation (4 cases): 
Administrative or procedural relationships such as ``Clinicians prescribe Opioid'' and ``Women on BUP had documented Opioid misuse.''

\item Methodological relationships (2 cases): Experimental 
procedures such as ``Anisomycin used to detect GNPTG.''

\item Specialized reporting contexts (1 case): Context-specific descriptions not representing biological relationships.
\end{enumerate}

These rejections demonstrate RELATE's ability to distinguish between 
legitimate biomedical relationships and other types of associations commonly found in scientific literature. All rejections were semantically distant from Biolink predicates, making the \texttt{NONE} decision appropriate.

\subsubsection{ChemProt Dataset Rejections}

Analysis of the ChemProt test set (695 cases with ground truth labels) revealed 18 instances where RELATE selected \texttt{NONE}. Of these, only 5 were correct rejections, yielding a rejection precision of 27.8\% (72\% false rejection rate among rejected cases).

The 13 incorrect rejections occurred despite appropriate candidate 
predicates being retrieved with moderate-to-high similarity scores during Stage 2 (Section~\ref{stages:stages2}). Examples include:

\begin{enumerate}
\item ``loperamide modifies corticotropin-releasing hormone action'' retrieved \textit{modulator} with similarity score 0.661, yet was rejected in Stage 3.

\item ``ammonium sulfate reduces CBG concentrations'' retrieved 
\textit{downregulator} with similarity score 0.632, yet was rejected in Stage 3.
\end{enumerate}

These patterns indicate that incorrect rejections stem from overly 
conservative contextual evaluation during LLM-based refinement in Stage 3 rather than retrieval failure in Stage 2. The system prioritizes precision 
by applying strict semantic criteria when assessing predicate-context fit, which reduces false positives but increases false rejections in ambiguous cases.

\subsection{Negation Handling}\label{apd:negation}
We conducted a manual evaluation of negation detection on the HEAL dataset. Of 2,411 relationship extractions from 2,400 abstracts, we identified 123 (5.1\%) negated relation texts through manual review.

RELATE correctly flagged 74 as negated assertions, yielding:
\begin{enumerate}
\item Precision: 96.1\% (74 true positives / 77 flagged as negated)
\item Recall: 60.2\% (74 true positives / 123 actual negations)
\item F1-score: 74.2\%
\end{enumerate}

The high precision indicates that when RELATE identifies a negation, it is almost always correct (only 3 false positives). The moderate recall indicates that approximately 40\% of negated statements are missed (49 false negatives), typically involving:

\begin{enumerate}
\item Implicit negations (e.g., ``failed to affect'', ``unable to deliver'')
\item Comparative negations (e.g., ``no significant difference'')
\item Complex syntactic constructions not captured by pattern matching
\end{enumerate}

The negation categories in the 123 manually identified cases were:
\begin{enumerate}
\item Explicit negations (``not'', ``no'', ``does not''): 90 cases (73\%)
\item Implicit negations (``failed to'', ``unable to''): 21 cases (17\%)
\item Comparative negations (``no significant difference''): 12 cases (10\%)
\end{enumerate}

RELATE's negation handling maintains conceptual accuracy by identifying the underlying relationship type while flagging the negation polarity, enabling accurate representation of negative findings in KGs.

\section{Predicate and Descriptor Distributions}\label{apd:data_descriptions}

ChemProt defines 10 classes for chemical-protein interactions: 9 positive relationship types and 1 negative class (\textit{NOT/no interaction}). We utilize all 9 positive relationship types, yielding 30 positive descriptors (Table~\ref{tab:top_predicates_distribution_chemprot}). ChemProt's NOT class, which indicates no interaction exists, is handled through RELATE's explicit rejection (\texttt{NONE}) mechanism rather than treating \textit{NOT/no interaction} as a mappable predicate. This design ensures that only valid biomedical relationships are standardized to ontology predicates while filtering non-relationships before KG integration.

After the preprocessing stage (Section~\ref{stages:stages1}), the ontology database stores both positive predicates and descriptors and their LLM-negated versions. For ChemProt, this yields 18 stored 
predicates (9 positive + 9 NEG variants) with 60 total descriptor texts (30 positive + 30 negated). For Biolink, this yields 568 stored predicates (284 positive + 284 NEG variants) with 2,208 total descriptor texts (1,104 positive + 1,104 negated).

Tables~\ref{tab:top_predicates_distribution_chemprot} and~\ref{tab:top_predicates_distribution_heal} show the descriptor distribution for the most frequent predicates. In ChemProt (Table~\ref{tab:top_predicates_distribution_chemprot}), \textit{substrate} and \textit{agonist} together account for over one-third of all positive descriptors (36.67\%). In Biolink (Table~\ref{tab:top_predicates_distribution_heal}), the top 10 predicates account for 41.30\% of all positive descriptors, with \textit{coexists with} (16.12\%) and \textit{related to} (8.70\%) being the most prevalent. The remaining 274 Biolink predicates account for 58.70\% of descriptors, demonstrating the diversity of biomedical relationship types captured in the ontology.

\begin{table*}[htbp]
\floatconts
  {tab:top_predicates_distribution_chemprot}
  {\caption{ChemProt Dataset. All 9 positive relationship predicates ordered by descriptor count. ChemProt's 10th class (NOT/no interaction) is handled via RELATE's rejection mechanism (\texttt{NONE} option, Stage 3) rather than predicate mapping.}}
  {\begin{tabular}{@{}lp{10cm}c@{}}
    \toprule
    \textbf{Predicate} & \textbf{Sample Descriptors} & \textbf{Descriptor count (\%)} \\
    \midrule
    substrate      & Product of, Substrate product of & 6 (10.00\%) \\[1mm]
    agonist        & Agonist-activator, Agonist-inhibitor & 5 (8.33\%) \\[1mm]
    upregulator    & Activator, Indirect upregulator & 4 (6.67\%) \\[1mm]
    downregulator  & Inhibitor, Indirect downregulator & 4 (6.67\%) \\[1mm]
    regulator      & Direct regulator, Indirect regulator & 3 (5.00\%) \\[1mm]
    part of        & Indicates that the chemical is a structural component or subunit of the protein complex & 2 (3.33\%) \\[1mm]
    antagonist     & Chemical substance that binds to and blocks the activation of certain receptors on cells, preventing a biological response & 2 (3.33\%) \\[1mm]
    modulator      & Modulator-activator, Modulator-inhibitor & 2 (3.33\%) \\[1mm]
    cofactor       & A non-protein chemical compound that is required for an enzyme's biological activity & 2 (3.33\%) \\
    \bottomrule
  \end{tabular}}
\end{table*}
\begin{table*}[htbp]
\floatconts
  {tab:top_predicates_distribution_heal}
  {\caption{Biolink Dataset. Top 10 of 284 predicates by descriptor count. Percentages calculated from 1,104 total positive descriptors across all 284 Biolink predicates used in this study.}}
  {\begin{tabular}{@{}lp{9cm}c@{}}
    \toprule
    \textbf{Predicate} & \textbf{Sample Descriptors} & \textbf{Descriptor count (\%)} \\
    \midrule
    coexists with           & holds between two entities that are co-located in the same aggregate object, process, or spatio-temporal region & 178 (8.06\%) \\[1mm] 
    related to              & holds between two entities & 96 (4.39\%) \\[1mm] 
    located in              & holds between a material entity and a material entity or site within which it is located (but of which it is not considered a part) & 40  (1.81\%) \\[1mm]
    part of                 & holds between parts and wholes (material entities or processes) & 34  (1.49\%) \\[1mm]
    temporally related to   & holds between two entities with a temporal relationship & 30  (1.36\%) \\[1mm]
    has part                & holds between wholes and their parts (material entities or processes) & 21  (0.95\%) \\[1mm]
    precedes                & holds between two processes, where one completes before the other begins & 15  (0.68\%) \\[1mm]
    causes                  & holds between two entities where the occurrence, existence, or activity of one causes the occurrence or generation of the other & 15 (0.68\%) \\[1mm]
    affects                 & Describes an entity that has an effect on the state or quality of another existing entity & 14  (0.63\%) \\[1mm]
    has output              & holds between a process and a continuant, where the continuant is an output of the process & 13 (0.63\%) \\[1mm] 
    \bottomrule
  \end{tabular}}
\end{table*}

\end{document}